On the Andromeda to Milky Way mass-ratio

Giovanni C. Baiesi Pillastrini[1]
*Sezione di ricerca Spettroscopia Astronomica – Unione Astrofili Italiani c/o IASF-INAF Via del Fosso del Cavaliere, 100, 00133 Rome, Italy*

Abstract

We have explored the hypothesis that the total mass-ratio of the two main galaxies of the Local Group: Andromeda Galaxy (M31) and the Milky Way (MW) can be constrained measuring the tidal force induced by the surrounding mass distribution, M31 included, on the MW. We argue that the total mass-ratio between the two groups can be approximated, at least qualitatively, finding the tidal radius where the internal binding force of the MW balances the external tidal force acting on it. Since M31 is the massive tidal "perturber" of the local environment, we have used a wide range of M31 to MW mass-ratio combinations to compute the corresponding tidal radii. Of them, only few match the distance of the zero-tidal shell i.e. the shell identified observationally by the outermost dwarf galaxies which do not show any sign of tidal effects. This is the key to constrain the best mass-ratio interval of the two galaxies. Our results favour a solution where the mass-ratio ranges from 2 to 3 implying a massive predominance of M31.

[1] *Permanent address: Via Pizzardi, 13 – 40138 Bologna – Italy; email: gcbp@it.packardbell.org*

1. Introduction

The total mass-ratio between the Andromeda Galaxy (M31) and the Milky Way (MW) is an intriguing puzzle. Very recent papers favour a mass-ratio close to unit suggesting that M31 is as massive as the MW (hereafter M31 and MW are intended as groups including their satellite dwarf galaxies). Comparing the H I rotation curve of M31 with the analogue of the MW obtained from trigonometric parallaxes and proper motions of masers in star formation regions, Reid et al. (2009) concluded that the dark matter halo of M31 and MW are comparably massive confirming previous suggestion of Evans et al. (2000). Furthermore, Evans & Wilkinson (2000) and Gottesman et al. (2002) claimed for a total mass of M31 lesser than that of the MW. On the other hand, Karachentsev et al. (2009) studying the peculiar velocity pattern around the Local Group, inferred a M31/MW mass-ratio of 1.25 evidencing a small mass predominance of M31 on the MW. This recent result confirms partially an older one based on timing arguments which found a larger mass-ratio of between 1.3 and 1.7 (Zaristky 1999). In the last decade, a great effort has been done to improve the mass evaluation of the two galaxies. The large amount of dwarf galaxies recently discovered even at large Galactocentric distances (Belokurov et al. 2006a,b, 2008; Zucker et al. 2006 a,b; Willman et al. 2005a,b; Irwin et al. 2007; Walsh et al. 2007) has been used to better modeling the structural properties of the halo deriving new estimations of the total mass of the Milky Way (Battaglia et al. 2006; Dehnen et al. 2006; Besla et al. 2007; Kalberla et al. 2007; Smith et al. 2007; Xue et al. 2008; Li & White (2008). Similar objects have been found around M31 (Zucker et al. 2004a, 2007; Martin et al. 2006; Majewski et al. 2007) which helped to improve the mass estimation of M31 (Majewski et al. 2007; Seigar et al. 2008). However, even if the new data have largely increased our knowledge of both M31 and the MW, there are not a general concordance between their estimated masses. For example, in a recent paper Xue et al. (2008) estimated the mass of the MW dark matter halo using a set of 2401 halo stars from the SDSS as kinematic tracers and assuming a NFW halo profile, they found a value of $\sim 1 \times 10^{12}$ $M_{sun}$ reopening the question of whether all of the MW satellite dwarf galaxies are on bound orbits. The new sample of discovered satellites is generally assumed bound within the MW dark halo, but the assumption seems to hold only if the mass is $\geq 2 \times 10^{12}$ $M_{sun}$ (Peebles 1995; Wilkinson & Evans (1999); Sakamoto et al. 2003; Loeb et al. 2005; Li & White 2008). It is noteworthy that almost all previously published estimations fall within the above mass interval (e.g. Klypin et al. 2002; Bellazzini 2004; Karachentsev 2005). More controversial is the case of M31 after the recent study of Seigar et al. (2008) that found a total mass of $0.8 \times 10^{12}$ $M_{sun}$ which confirms similar low-mass estimations (e.g. Evans & Wilkinson 2000; Gottesman et al. (2002); Klypin et al. 2002; Karachentsev 2005; Majewski et al. 2007) but in contrast with other estimations $\geq 3 \times 10^{12}$ $M_{sun}$ (Peebles 1996; Loeb et al. 2005). From such mass intervals a very wide range of possible M31 to MW mass-ratios can be assumed. Can this issue be disentangled using a different approach? Karachentsev (2005) suggested a strategy based on the tidal relationships among gravitationally interacting bodies, e.g. the tidal interactions between a dominant galaxy and its satellites. However, this method may underestimate significantly the total amplitude of the tidal force since it does not take into account further tidal influences (even if small) coming from the external surrounding mass distribution. On the contrary, in a similar but statistical approach, Baiesi Pillastrini (2006) took into account the gravitational potential induced by extended environments to estimate the tidal fields acting on 11 galaxy groups used as test particles. Their total masses have been established by examining the tidal limits set by the surrounding mass distributions on these groups. In the present work, instead of attempting to constrain the total mass of the Milky Way (thought as a group), since M31 is the major tidal "perturber" of the local environment, the best M31 to Mw mass-ratio we have been identified using a wide range of combinations of mass-ratio to compute the corresponding *tidal radii* around the MW. Of them, only few will match the distance of the *zero-tidal shell* i.e. the shell where all forces cancel out each other. The location of this shell will be identified observationally looking at the physical properties of the outermost dwarf galaxies surrounding the MW. They should *not* show any sign of observational effects of tidal stripping like mass (stars) or gas (H I) loss, streaming tail, irregular morphology and so on. This is the key to constrain the best interval of M31 to MW mass-ratios and the corresponding total masses of each system. To disentangle this issue, in Section 2 we present the method based on the tidal theory. The application of the method is performed in Section 3. Then, in Section 4 we discuss our results. Finally, in Section 5 the concluding remarks.

2. The tidal approximation

Following Baiesi Pillastrini (2006), the strategy involves approximate descriptions of external influences incorporating the larger external influence through static tidal field estimated on the basis of the present-day locations of the nearby objects enclosed in a spherical volume representative of the external density distribution. In other words, we assume that the source of the tidal force is due to a time-independent gravitational potential generated by the "point mass" distribution of the surrounding galaxies and galaxy groups centred on the MW frame of reference. We assume the tidal effect as a static tidal limitation spatially fixed by the *tidal radius* beyond which the binding force dominates the internal dynamics of the MW, while external objects would be torn apart by the tidal field. Then, the tidal force acting on the MW and its satellites can be expressed by

$$F_{tidal,a} \equiv -\frac{d^2\Phi_{ext}}{dR_a dR_b} R_b \equiv F_{ab} R_b \quad (1)$$

where $\Phi_{ext}$ is the external potential and $R$ is the radius vector in the MW reference frame. Then, if the MW is subjects to the action of $N$ nearby galaxy groups and galaxies at a position vector $r_g$ and mass $m_g$, the external potential is given by

$$\Phi_{ext} = -G\sum_N \frac{m_g}{|r_g|} \quad (2)$$

and, the tidal tensor is

$$F_{ab} = \sum_N \left(\frac{m_g}{|r_g|^3}\right)\left(\frac{3(r_g)_a (r_g)_b}{|r_g|^2} - \delta_{ab}\right) \quad (3)$$

where the gravitational constant G=1 and $\delta_{ab}$ is the Kronecker delta. It follows that the amplitude of the tidal force is

$$F_{tidal} = |F_{aa} R_a| \quad (4)$$

where $F_{aa}$ are the three eigenvalues corresponding to the principal axes of the 3 x 3 symmetric matrix $F_{ab}$. By assuming that the MW and its satellites is a group approximately spherically symmetric and dynamically relaxed, the condition $F_{tidal} = F_{binding}$ must be satisfied. Plugging in

$$F_{tidal} = \frac{M}{R^2} \quad (5)$$

where $M$ and $R$ are the fiducial virial mass and radius of the MW.
Then, the tidal radius is

$$R_t = \left(\frac{M}{F_{tidal}}\right)^{\frac{1}{2}} \quad (6)$$

3. Application

3.1. Methodology

The application has been organized in the following way: i) we calculate the net tidal force acting on the MW assuming that it is induced by the local environment enclosed in a spherical volume of 5 Mpc-radius as the first approximation of the sampling; ii) we know that ~ 80 per cent of the tidal amplitude is generated by M31, the nearest and massive companion. Keeping fixed the remaining 20 per cent due to the farthest masses and running a grid of mass parametrization for M31, entering in Eq.(3), and MW, in Eq.(6), as a function of a wide range of combinations of mass-ratios, we obtain the corresponding range of the computed tidal radii $R_t$. Knowing that the mass estimations of the MW ranges from 1 to 2.5x10$^{12}$ M$_{sun}$, while those of M31 ranges from 1 to 3.5x10$^{12}$ M$_{sun}$, the mass parametrization for M31 is 1, 1.5, 2, 2.5, 3, 3.5 (x10$^{12}$ M$_{sun}$) and 1, 1.5, 2, 2.5 for the MW. Then, the grid of 6 x 4 = 24 combinations of mass-ratios has been run to obtain the corresponding 24 tidal radii $R_t$; iii) the $R_t$ (one or many) that best matches R$_{ZTS}$ (i.e. the distance of the zero-tidal shell discussed in Section 1) enables to constrain the best M31 to MW mass-ratio.

3.2. Data

To perform our analysis we have used the data collected by Pasetto & Chiosi (2007) in order to study the planar distribution of galaxies in the Local group. From their table 2, the Galactic coordinates, distances and mass estimations of 6 massive galaxy groups located approximately within 5 Mpc-radius from the MW have been acquired. Similarly, from their table 3, a supplementary list of 22 galaxies lying in the same volume have been added to our database in order to have a sample representative of the real mass distribution surrounding the MW. A detailed description of the data (references, corrections and uncertainties) can be found in their Section 4. If the light is a tracer of the mass, these objects represent almost all light within the sampled sphere.

3.3. Simplifying assumptions

We have identified the location of $R_{ZTS}$ on the basis of the physical properties of the dwarf galaxy satellites lying at increasing distances from the MW. For instance, Leo V, a small dwarf galaxy discovered at a distance of ~180 Kpc, shows clear signs of strong tidal stripping (Walker et al. 2009). Another example of apparent tidal effect has been found on the dwarf spheroid Leo I which resides at a Galactocentric distance of ~250 Kpc (Sohn et al. 2007; Muñoz et al. 2008) even if the tidal origin of this effect has been recently challenged (Penarrubia et al. 2009). Besides, a recent analysis on the H I content of the MW satellites shows that almost all satellites within a radius of ~ 270 Kpc are undetected in H I (Grcevich & Putman, 2009). Even if a gas loss mechanism due to ram pressure stripping is preferred to a tidal stripping origin, such H I depletion suggests that these objects are likely in bound orbits within the MW halo potential. Therefore, it seems to us that inside ~300 Kpc-radius, the tidal influence of the MW on the structures of the satellite population is evident even if the lack of gas in such small dwarfs may be attributed to other physical phenomena of sweep-out as stellar winds and supernova shell-bursts. Therefore, it is not so clear that either the presence or lack of gas and interstellar dust can be used to discriminate objects subject to tidal influence, especially after the cases of NGC 185 and 205. They are two elliptical dwarf galaxies, both satellites near to M31, which show abundance of HI and dust contents in contrast to the clear signs of gravitational tidal interaction visible in their structures (Young & Lo, 1997). However, we can reasonably assume that the combination of finding a relevant HI (and/or dust) content within an undisturbed morphology is indicative of negligible tidal influence. This could be the case for two dwarf galaxies at ~400 Kpc from the MW detected by their H I content: Leo T and Phoenix. Even if they have been added (singly or together) to satellite lists of the MW (Karachentsev 2005; Simon & Geha, 2007; Madau et al. 2008; Grcevich & Putman, 2009), their structures do not seem to be influenced by the Galactic tidal field showing a relevant gas content with no signs of tidal stripping or ram pressure (Irwin et al. 2007; Simon & Geha, 2007; Young et al. 2007). Besides, they are both located at comparable distances from the MW and M31. If this is the case, Leo T and Phoenix lie at the boundary of the MW sphere of influence and their Galactocentric distances of ~400 Kpc is assumed as our fiducial $R_{ZTS}$ allowing an error of ± 50 Kpc. As already stated, this is a very simplify assumption based on few observed features of only two objects. Besides, we do not know nothing about their orbits, in particular if they are bound to the MW. There are not proper motion measurements to establish it; we can only suppose that they are probably observed at the apogalacticon where dynamical friction and tidal effect would be negligible. In such a case, as well as in the unbound one, $R_{ZTS}$ may turn out overestimated increasing dramatically the mass ratio in favour of M31. It is worthy to note that our $R_{ZTS}$ is smaller than ~700 Kpc inferred by Karachentsev (2005) and slightly larger than ~300 Kpc reported by Lin et al. (1995) but matches the predicted limit within which is expected to find the "missing" ultra-faint dwarf population (Diemand et al. 2007; Tollerud et al. 2008; Koposov et al. 2009).
Finally, we assume 280 Kpc as the fiducial virial radius of the MW (Xue et al. 2008; Shattow & Loeb 2009).

3.4. Error in estimating the tidal force

The major source of uncertainty of $F_{tidal}$ can be due to the assumed mass estimations of the sampled objects. Cen (1997) found that cluster virial mass estimations are, on average, 20 per cent underestimated with respect to the simulations. A result which has been confirmed by Evans et al. (2003) that, on the basis of 10,000 Monte Carlo simulations, demonstrated that at least 87 per cent of the virial mass estimations of galaxy groups are below the true mass. If our data are affected in likewise manner, we expect an overestimation of ~ 8-10 per cent on the calculation of $R_t$. As will be discussed later, such a percentage *could* change, even qualitatively, our result. Note that the error on $R_t$ is significantly reduced by the ½ exponent of Eq.(6).

4. Results and discussion

As can be seen in Fig. 1, the expected decreasing sequence of the tidal radii $R_t$ from low to high mass-ratios, intersects $R_{ZTS}$ in correspondence of the M31/MW mass-ratio = 3. Allowing an error of ± 50 Kpc, the mass-ratio interval that best matches $R_{ZTS}$ ranges from 2.5 to 3.5.

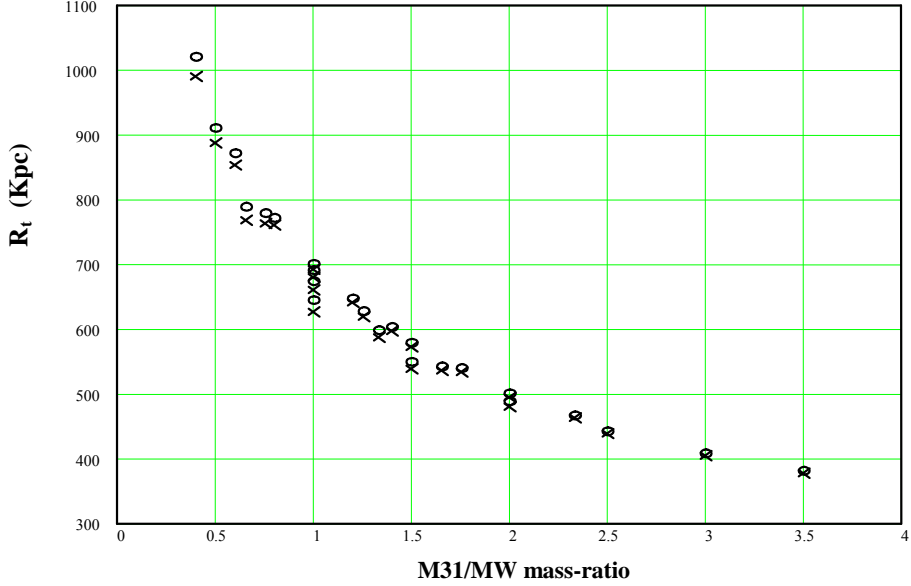

**Figure 1.** Plot of the tidal radii computed by Eq.(6) as a function of 24 combinations of the M31 to MW mass-ratio. Dots are the tidal radii computed using the mass distribution within the sampled sphere of 5 Mpc-radius. Crosses are those computed in the expanded sphere of 20 Mpc-radius (see text).

This is a straightforward demonstration that M31 is more massive than the MW and, even if the uncertainties are unknown, this result can be considered quite reliable, at least qualitatively. In fact, one may put suspects on the reliability of the computed value of $F_{tidal}$. As stated before, the tidal force is fairly determined when its cumulative amplitude converges asymptotically within the sampled sphere. To test it, the amplitude of the tidal force has been computed for a set of concentric spheres of increasing 2 Mpc-radius from the MW. We have found that the development of the cumulative amplitude of the tide tends to converge asymptotically but *not* completely indicating that the boundary of the sampled sphere is not large enough to incorporate the major share of the gravitational influence. This means that the asymptote will be approached farther away at larger distances so that the tidal force turns out underestimated and $R_t$ overestimated (even if moderately).

Therefore, it becomes very important to quantify how large is the error on $R_t$ due to insufficient sampling of the surrounding mass distribution. We proceed expanding the sampled sphere up to 20 Mpc-radius in order to include in the calculation the masses of the Virgo cluster and the most relevant galaxy groups within it. From table 1 of the UZC-SSRS2 group catalog (Ramella et al. 2002) we extract Galactic coordinates, radial velocities and virial masses of the following objects: U478, U480, U490 (Virgo), S129 and S190 (for simplicity, the distances have been derived from the well-known Hubble relation that is, $V_r/H_o$ where $V_r$ is the radial velocity and $H_o$ is the Hubble constant assumed of 70 $Kms^{-1}Mpc^{-1}$).

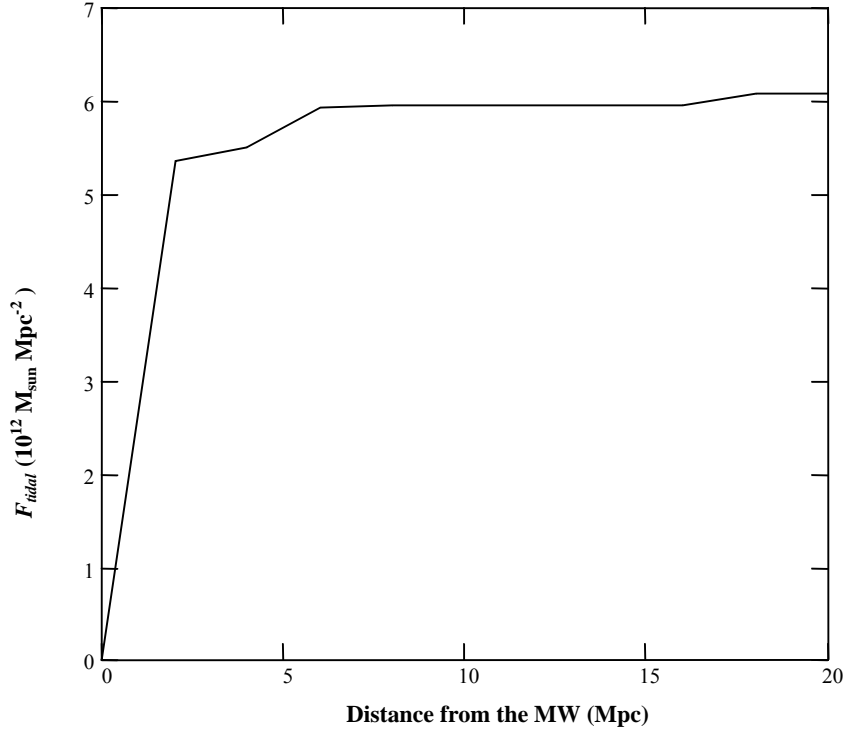

**Figure 2.** Plot of the cumulative amplitudes of $F_{tidal}$ as a function of increasing distance of 2 Mpc-bin from the MW.

In Fig.2 the cumulative amplitude of $F_{tidal}$ is apparent. As expected, it increases sharply within the first 2 Mpc-bin due to the presence of the massive M31 galaxy; **the** bump at 6 Mpc-radius is due to the rich group U480; from 6 to 16 Mpc-radius it increases very slowly then, a very small bump at the distance of Virgo followed by a flat line indicates that the mass of the Virgo cluster provides the largest tidal influence on the MW. However, as can be seen in Fig. 1, the small increment of $F_{tidal}$ found inside 20 Mpc-radius provides a negligible ~ 1 percent decrement of $R_t$ at high mass-ratios and ~ 2-3 percent at the low ones which do not change qualitatively our result. More serious could be the error affecting $F_{tidal}$ due to systematic underestimations of the virial masses discussed in Section 3.4. If one would take into account such a bias, the values of $R_t$ should be rescaled down of ~10 per cent intersecting $R_{ZTS}$ at a lower mass-ratio = 2.5 allowing a re-evaluated mass-ratio interval ranging from 2 to 3. Therefore, assuming a conservative point of view, we assume this new result as our fiducial one. Finally, from the fiducial mass-ratios we can evaluate the corresponding (fiducial) mass interval for both M31 and the MW having in mind the initial constraints derived from the published set of mass estimations. Roughly, the total mass of the Milky Way would range from 1 to $1.5 \times 10^{12}$ $M_{sun}$, while Andromeda Galaxy between 2 and $3 \times 10^{12}$ $M_{sun}$.

5. Concluding remarks

The main accomplishment of this paper is the introduction of a method based on the tidal theory in order to study the total mass-ratio between the two dominant galaxies of the Local Group: M31 (Andromeda Galaxy) and the Milky Way. The mass-ratio between the two galaxy groups has been established by examining the tidal limits set by the surrounding mass distribution on the Milky Way and comparing them with the distance at which the outermost dwarf galaxies do not show any apparent effects of tidal stripping. We have demonstrated, at least qualitatively that the Andromeda Galaxy is more massive than the Milky Way by a factor between 2 and 3. The recent finding by Reid et al. (2009) predicting equivalent total masses for M31 and the MW is clearly in contrast with our result. As can be seen in Fig. 1, a unit mass-ratio would be satisfied by tidal radii ranging from ~600 to 700 Kpc, very close to the outskirt of M31! If our result is correct, the Andromeda Galaxy is likely embedded in a larger and more massive dark halo than that of the Milky Way. It seems to us that these discrepant results are characterized by two different methodologies used to determine the total mass of the MW and M31. One investigates the physical properties considering the objects as separate systems, while the other analyze the gravitational interactions in the context of the local environment. The former leads to a lower mass estimation for M31 probably because the current constraints on the shape and extent of the dark matter halos are still doubtful (the missing faint-dwarf satellite problem) and model dependent. The latter suffers of large uncertainties in the distances and mass determinations of the sampled objects as in the present work. In any case, one should take into account that the reliability of our result is weakened by the unknown uncertainties affecting the assumed parameters ($R_{ZTS}$ for instance) and kinematical data and, in spite of the very conservative behaviour on the evaluation of the result, the lacking of a detailed error analysis prevents its acceptance from a quantitative point of view. This is the true limit of our analysis which prevents a deeper study of many related problems connected with the tidal interactions among the MW, M31 and neighbouring galaxies and groups. For example, an interesting question arise from the physical meaning of $R_{ZTS}$ : is it coincident with the dark halo radius or it lies beyond? The question is not trivial. If coincident, the larger mass of M31 implies a larger tidal influence on the MW. Consequently, because of proximity of the two galaxies, it would follow that the M31 halo should overlap or encompass the MW one. The answer to this question will be possible only by obtaining proper motion determinations for the outer satellite orbits allowing to understand if these extreme objects are bound and were already shaped by past central encounters with the host galaxy. And, finally, providing constraints on the profile of the dark matter distribution out of the virial radius as well as the $R_{ZTS}$.

Aknowledgements

We thank the referee, Stephen T. Gottesman, for useful criticisms regarding the reliability of the assumed $R_{ZTS}$ and comments that greatly improved the paper.